\documentclass[journal]{IEEEtran}
\usepackage{multirow}

\ifCLASSINFOpdf
 \usepackage[pdftex]{graphicx}
\usepackage{subcaption}
\usepackage{xcolor}
\usepackage{cite}
\else
\fi
\begin{document}
%
\title{TMS-Crossbars with Tactile Sensing}
%
%
%

\author{R. Chithra, A.R. Aswani, and A.P. James
}

%
%

\markboth{}%
{}
%



\maketitle

\begin{abstract}
The first stage of tactile sensing is data acquisition using tactile sensors and the sensed data is transmitted to the central unit for neuromorphic computing. The memristive crossbars were proposed to use as synapses in neuromorphic computing but device intelligence at the sensor level are not investigated in literature. We propose the concept of Transistor Memristor Sensor (TMS)-crossbar by including sensor to memristor crossbar array configuration in the input layer of the neural network architecture. 2 possible cell configurations of TMS crossbar arrays: 1 Transistor 1 Memristor 1 Sensor (1T1M1S) and 2 Transistor 1 Memristor 1 Sensor (2T1M1S) are presented. We verified the proposed TMS-crossbar in the practical design of analog neural networks based Braille character recognition system. The proposed design is verified with SPICE simulations using circuit equivalent of FLX-A501 force sensor, TiO$_2$ memristors and low power 22nm high-k CMOS transistors. The proposed analog neuromorphic computing system presents a scalable solution and is possible to encode 125 symbols with good accuracy in comparison with other Braille character recognition systems in the literature. The benefits of analog implementation of the TMS crossbar arrays is substantiated with results of accuracy, area and power requirements in comparison with the binary counterparts.
\end{abstract}
\vspace{-4pt}

%
\IEEEpeerreviewmaketitle
\vspace{-10pt}
\section{Introduction}
\IEEEPARstart{T}{} he "sense-of-touch" or tactile sensing involves sensor arrays that enable a specific surface area to be monitored \cite{sym0008}. It is estimated that for electronic Skin (eSkin) applications the fingertip sites itself require approximately 15 $\times$ 10 sensing arrays  \cite{article201904765, citeee123}. The sensed data generated by these receptors are transmitted via large number of cables or wireless modules for neural computing which will add to the weight and complexity. The local computations at receptor sites is energy efficient and reduces complexity. However, tactile sensing at edge is a topic that is less explored. Here, we propose an intelligent edge-AI computing solution to process the sensed data at the sensor level in analog domain. The concept of Transistor-Memristor-Sensor (TMS) crossbar array and its neural computations in analog crossbar circuits is proposed in this paper. The memristive crossbar circuits are generally implemented with 1 Transistor 1 Memristor (1T1M) crossbar arrays \cite{8667457}. However, efficient integration of analog sensor units next to memristive neural network is an open problem for tactile sensing.  

{In this paper, we propose the concept of Transistor-Memristor-Sensor (TMS) crossbar neural computing. The possible cell configurations for TMS crossbar array, (a) 1 Transistor 1 Memristor 1 Sensor (1T1M1S) and (b) 2 Transistor 1 Memristor 1 Sensor (2T1M1S), is presented. The performance and robustness of the proposed method is evaluated using the Braille character recognizing system. The proposed system consists of a tactile sensory patch with analog neural computing capabilities. The tactile sensory patch is implemented using the TMS crossbar arrays which forms the first layer of neural network. Till date, no circuit level implementations of Braille character recognition problem using neural networks has been reported in the literature. Most of the works in braille technology using neural networks converts images or binary encoding of braille system into corresponding characters \cite{sym12071069, sym0007, 9152576, sym0001}. However, such systems are not scalable in real-time implementation due to similar combination of dots, making the system design complex. We propose a scalable hardware system with an analog design combined with neural network computation. The proposed system gives better performance regardless of the number of characters added on it. The accuracy of the proposed system is evaluated in varying noise conditions. The advantage in power and area of the proposed analog TMS-based system in implementing the braille technology is shown with respect to the binary implementations.}  

This paper is organized into following sections: Section II introduces the concept of TMS crossbars with 1T1M1S and 2T1M1S cell configurations. Section III describes how the TMS crossbar neural network can be implemented. Section IV reports the results and related discussions. Finally, Section V provides the concluding remarks.

%
%
%
%
\vspace{-9pt}
\section{TMS-Crossbars}
\subsection{1T1M1S crossbar configuration}

\begin{figure}
\begin{center}
  \includegraphics[width=60mm]{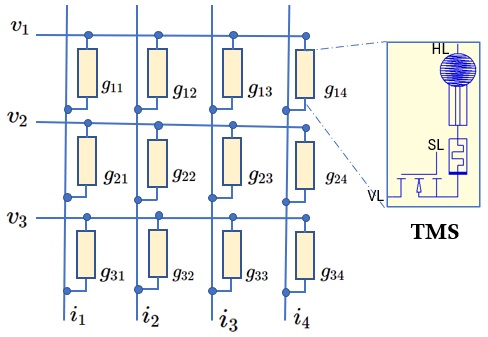}
  \caption{Transistor-Memristor-Sensor (TMS) crossbar array}
  \label{figure1}
  \end{center}
  \end{figure}
  
Fig. \ref{figure1} shows a single layer $m \times n$ neural network representation of sensor crossbar array with 1T1M1S cell configuration. The 1T1M1S configuration consists of a Transistor Selector, a memristor and a Sensor is shown in Fig. \ref{figure1}. The network consists of $m$ Horizontal Lines (HLs) and $n$ Vertical Lines (VLs); in Fig. \ref{figure1} $m=3$ and $n=4$. The inputs $ \left ( v_{1}, v_{2}, v_{3} \right ) $ are the supply voltage to the sensors in each HLs. The sensor will sense the external stimuli and convert it to corresponding electrical signals. The sensor output is then weighted using the memristive block. Similar to many prior arts, the transistor is used to control the accessibility of devices in a cell. Through the HLs and VLs, the synaptic weighting functions are executed in parallel. The output currents $ \left ( i_{1}, i_{2}, i_{3}, i_{4} \right ) $, are the weighted summation of inputs and can be read from the VLs. 

The effective conductance of each cell $g_{kl}$, with $k\in \{1,m\}, l\in \{1,n\} $, is the combined conductance of sensor, memristor and switching blocks. Thus the effective conductance of each cell becomes $g_{kl}=g_{S_{kl}}||g_{M_{kl}}||g_{T}$ where $g_{S_{kl}}, g_{M_{kl}}$ and $g_{T}$ are the effective conductance of sensor block, memristor and transistor at ON state respectively. When transistor is at OFF state, $g_{T}=0$ and hence $g_{kl}=0$. The transistor can be turned ON/OFF using the Select Line (SL). The total current in each column is the sum of individual currents flowing through each cell in that particular column. 
\vspace{-10pt}
\subsection{2T1M1S crossbar configuration}
\begin{figure}
	\centering
	\subfloat[]{\includegraphics[height=1.7in,width=1.4in]{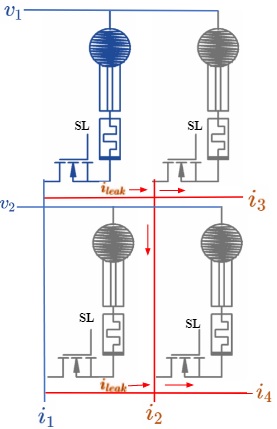}}\quad
	\subfloat[]{\includegraphics[height=1.7in, width=1.6in]{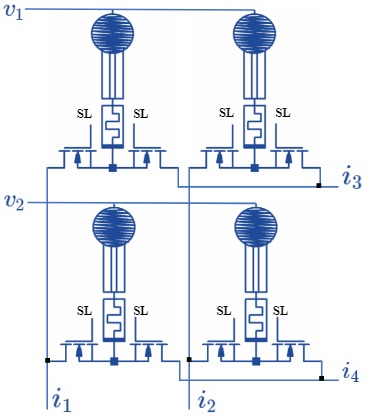}}\\	
	\caption{\small{(a) Current leakage in 1T1M1S cell configuration (b) 2-Transistor 1-Memristor 1-Sensor (2T1M1S) crossbar arrays}}
	\label{figure2}
\end{figure}

The VL readouts can provide only limited information regarding the sensed data for tactile applications. To increase the scalibility of system under implementation, more features need to extracted from the sensed data. Hence the HL read-out are also used in addition to VL readout. The weighted summation are taken along both VL and HL lines simultaneously. This increases the number of post-neurons in the neural network layer. The above mentioned 1T1M1S configuration can be used for VL-HL readout crossbar circuit. However, this results in sneak path current issues which degrades the detection accuracy \cite{GUL20191091}. The 1T1M1S configuration provides cell selection only in the VL lines of the crossbar. This will not ensure proper cell selection in the HL lines and hence results in leakage currents. Fig. \ref{figure2}(a) shows the leakage currents for VL-HL readouts in $2\times 2$ crossbars when $S_{11}$ is active, where $S_{11}$ denotes the sensor in first row and first column. Ideally the sensed data can be read from $i_{1}, i_{3}$ whereas $i_{2}, i_{4}$ should be zero. Since all nodes are short circuited, there will be current leakage and $i_{1}, i_{2}, i_{3}$ and $i_{4}$ will have the same currents. Thus, the core operation of a neuromorphic system, i.e, multiply and accumulate (MAC), cannot be implemented with the 1T1M1S crossbar array for VL-HL readouts. 

{For the VL-HL readouts in TMS crossbars, we propose 2 Transistors 1 Memristor 1 Sensor (2T1M1S) cell configuration as shown in Fig. \ref{figure2}(b). In 2T1M1S, the two transistor configuration is used to read the sensor data along the horizontal and vertical lines of the crossbars. A two transistors one memristor (2T1M) cell configuration was proposed in \cite{7010034, 7527510} for online learning of memristor where each memristor is connected to two inputs, one is the original input signal and other is the inverted input signal, obtained with the two transistors configuration. In contrast for the proposed 2T1M1S, one transistor act as a select switch in the VL$_{k}$ line and the other transistor act as switch in the HL$_{l}$ line. The output currents in VL$_{k}$ and HL$_{l}$ lines are weighted summation of inputs and nodal conductance.  
}

\subsubsection*{Negative Weights} 
The negative weights are implemented using two columns arrays with one column for storing positive weights and other for negative weights \cite{8471281}. The activation and deactivation of the transistor can be done using select lines (SL). Therefore the new node conductance $g'$ (or weights) will be difference in columns conductance (i.e. $ g'_{kl} \propto \left ( g_{kl}^{+} - g_{kl}^{-} \right ) $).


\section{TMS crossbar Neural Network Implementation}
The neural computing capabilities of TMS-crossbar arrays are explained with the braille character recognizing system. The proposed system consists of a tactile sensory patch with analog neural computing capabilities. The TMS-crossbar based sensory patch is capable of sensing inputs from the user and then calculate the weighted summation of sensed data. Here the sensor patch forms the first layer of neural network model and the subsequent layers are memristor crossbar array with 1T1M configuration \cite{8667457}. 
\vspace{-9pt}
\subsection{Sensor Crossbar array}
\begin{figure}
	\centering
	\subfloat[]{\includegraphics[height=1in,width=.8in]{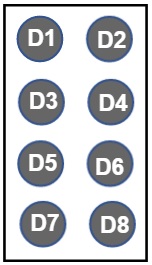}}\quad
	\subfloat[]{\includegraphics[height=1in, width=0.8in]{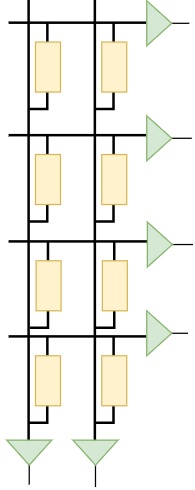}}	
	\caption{\small{(a) Dot Braille Character Structure (b) Neural Network Implementation of TMS crossbars}}
	\label{figure3}
\end{figure}

{In Braille system, each character is represented by 6 dots (D1, D2, D3, D4, D5, D6), Fig. \ref{figure3}(a). Since some of the braille symbols have the same notations, we use 2 additional dots (D7, D8) to differentiate them. Each dots represents one cell of TMS crossbar array in the 2T1M1S configuration.} The weighted summation of each dots are taken through HL and VL lines, Fig. \ref{figure3}(b). The first layer of our neural network consists of 4 preneurons and 6 postneurons. Of the 6 post neuron 2 neurons represents column out and 4 neurons represents row out in the crossbar array as shown in Fig. \ref{figure3}(b).

\begin{figure}
	\centering
	\subfloat[]{\includegraphics[height=1.2in, width=0.9in]{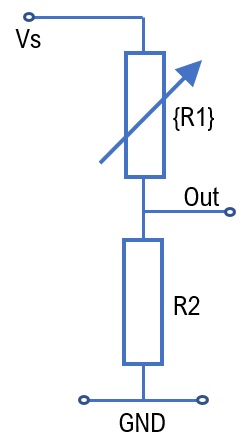}}\quad
	\subfloat[]{\includegraphics[height=1.2in, width=2in]{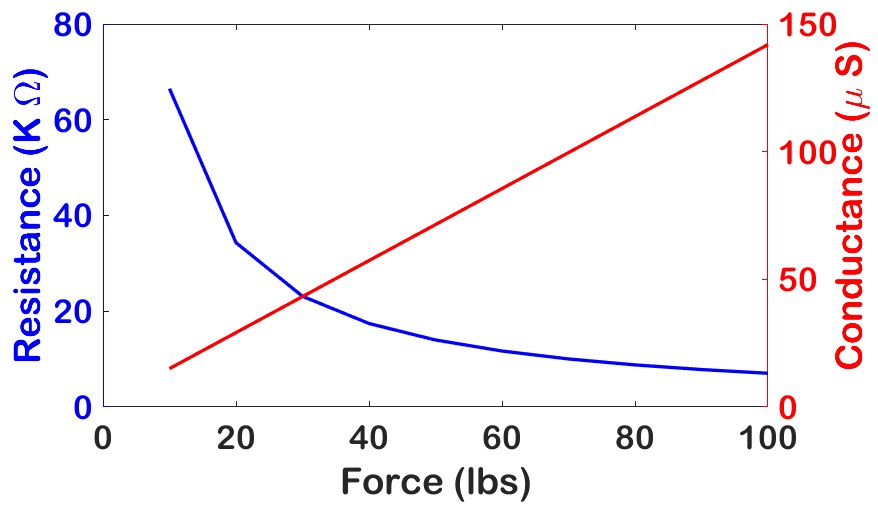}}\\	
	\caption{\small{(a) Equivalent circuit diagram of FLX-A505 force sensor (b) Resistance and Conductance variation with applied force of FLX-A505 force sensor}}
	\label{figure4}
\end{figure}

In the sensor block of 2T1M1S sensor crossbar array, the commercially available force sensing resistors (FSR) is used as unit elements for the Braille character inputs. The equivalent circuit model of the FSR, specifically Tekscan (Flexiforce sensors) \cite{flxa505}, used in the proposed design is presented in Fig. \ref{figure4}(a). It consists of a variable resistance part which can vary the resistance according to the applied force as per the FLX-A505 datasheet values \cite{flxa505}. The force dependent variable resistance equation is {$R_{1}(f)= 1/((1.5 \mu *f)+c)$}. Here $f$ is the applied force and the constant $c$ gives the bias voltage. The output is calculated using the voltage divider rule. Thus, with increase in force, resistance reduces and output voltage increases. The supply voltage to the sensor is 0.5V. The resistance variation along with the corresponding conductance variation with respect to applied force is plotted in Fig. \ref{figure4}(b). The force taken as a reference for detection is 20 lb.

 
\vspace{-9pt}
\subsection{Memristor Crossbar Design}
Except first layer, other layers are implemented with memristive crossbar arrays using 1T1M cell configuration \cite{dai2017forming,8667457}. The braille system designed using one-hot encoding with the memristor crossbar arrays requires weight updating with a training algorithm. The training is done using custom made Python programs. The trained weights are programmed to the corresponding memristors in the crossbar array. The hidden layer have ReLU activation function and the output neuron uses softmax activation function to get one-hot encoded output. The voltage signal applied to the HL lines is multiplied by the memristor conductance. The currents are then converted by means of the summation amplifier. 
\vspace{-9pt}
\subsection{Implementation of Sofmax Activation function}
\begin{figure}
	\centering
	\subfloat[]{\includegraphics[height=1.5in,width=1.3in]{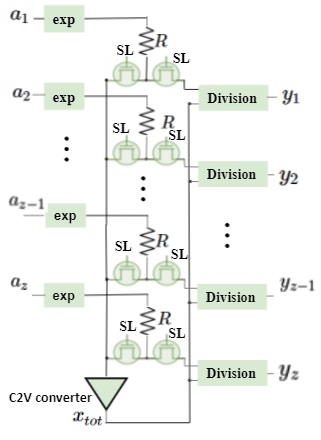}}\quad
	\subfloat[]{\includegraphics[height=1.4in, width=1.8in]{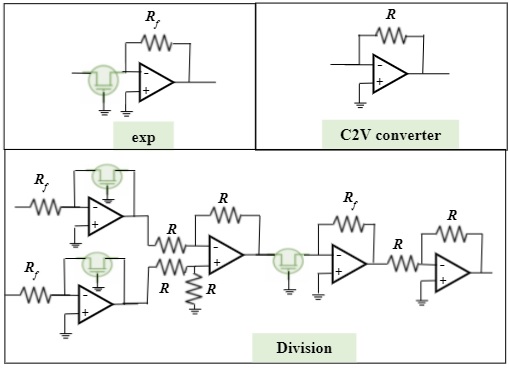}}	
	\caption{\small{(a) Equivalent circuit model of Softmax activation function. (b) Circuit diagrams of exp block, Current-to-Voltage converter and Division circuit.}}
	\label{softmax123}
\end{figure}

The onehot encoding of the braille system is done using softmax activation function at the output layer. The circuit level implementation of the softmax is shown in Fig. \ref{softmax123}. It consists of three main blocks: (a) exponential block, (2) Summation, and (3) Division block  \cite{patent123}. The novelty of our design is the softmax implementation using transistors amplifiers designed using 22nm COMS technology. The design is compatible with the preceding neural network layers designed using nanometer scale devices. The output of exponential block is $x_{z}=R_{f}I_{s} \exp\left( {\frac{a_{z}}{V_{T}}} \right) $ \cite{book11}. Where $V_{T}$ is the thermal voltage and $I_{s}$ is the saturation current. The summation block sum up the values from all the exponential block. {The conversion of column current to voltage is done with a 3 stage amplifier proposed in \cite{9233468} using 22nm CMOS node}. The output of the weighted summer will be $x_{tot}=\sum_{z} \left( \frac{R_{f}}{R} \right) x_{z}$. If $R=R_{f}$, $x_{tot}=\sum_{z} x_{z}$, i.e. summation of exponential operation. Substituting $x_{z}$, $x_{tot}=\sum_{z} R_{f}I_{s} \exp\left( {\frac{a_{z}}{V_{T}}} \right)$. Next process is to divide $x_{z}$ with $x_{tot}$. Fig. \ref{softmax123} shows the analog division circuit whose output is proportional to the division of two input values. We can divide two numbers by first calculating their logarithm, then calculate the difference between the log values and calculate the exponent of the difference value. The output of division circuit is $ R_{f}I_{s} \left( {\frac{V_{1}}{V_{2}}} \right) $ \cite{book11}. Substituting $x_{z}$ and $x_{tot}$ as $V_{1}$ and $V_{2}$, the theoretical value of $y_{z,theory}$ can be calculated as \cite{patent123}:  
\vspace{-9pt}
\begin{equation}
	y_{z,theory}= R_{f}I_{s} \left( {\frac{\exp \left( \frac {a_{z}}{V_{T}} \right) }{\sum_{i} \exp \left(  \frac{a_{z}}{V_{T}} \right) }} \right) 
	\label{theoretical}
\end{equation}






\vspace{-9pt}
\section{Results and Discussion}
The circuit design of TMS crossbar circuits and analog computation of Braille system were evaluated using SPICE models of realistic devices. The architecture level simulations for weight training, calculating performance accuracy with variation of noise were performed with Python programs. For 2T1M1S cell configurations, we use FLX-A505 with the characteristics as shown in Fig. 4 is used. {The sensed values can be captured for Braille character identification using the proposed TMS-crossbar array with current leakage of approximately 16\% from the ideal case (the maximum value with the presented hardware circuits)}. 22nm PTM high-k PTM models are used as transistor switches \cite{ptm123}. Memristor model was set with $R_{OFF}/R_{ON}=100K\Omega/1K\Omega$ \cite{9437294}. 
 \vspace{-9pt}
\subsection{Sensor Crossbar Array}
\begin{figure}
	\centering
	\subfloat[]{\includegraphics[scale=0.29]{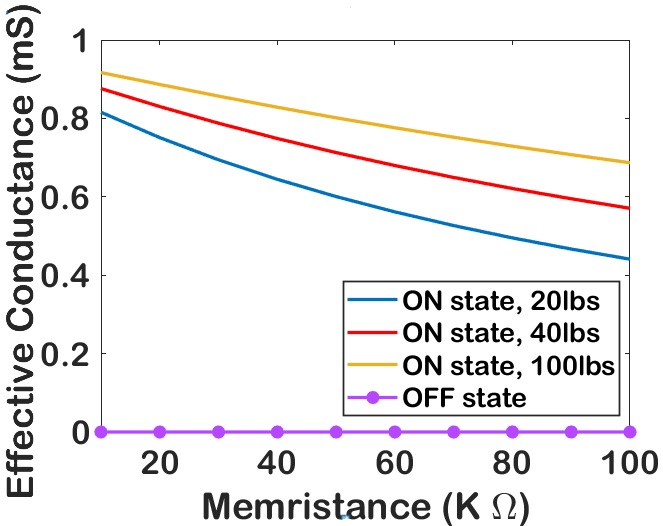} }  \quad
	\subfloat[]{\includegraphics[scale=0.29]{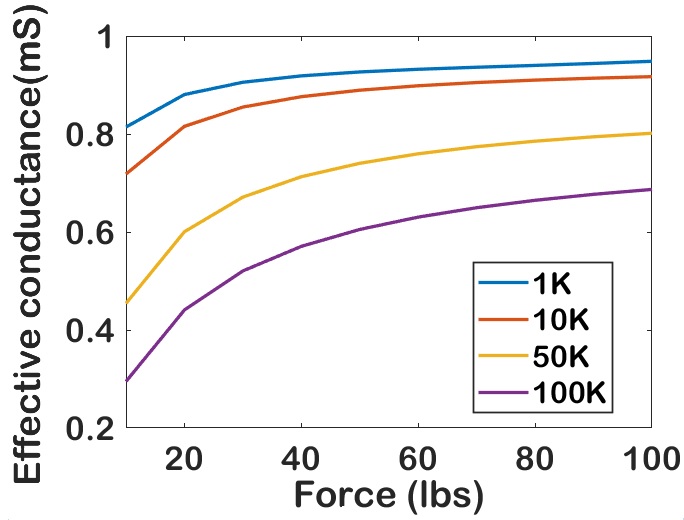}}\\	
	\caption{\small{Effective Conductance analysis of TMS crossbar arrays (a) Variation with respect to memristance (b) Variation with respect to applied force on FLX-A505 sensor}}
	\label{figure7}
\end{figure}

Fig \ref{figure7}(a) compares the effective cell conductance of a single cell $g$ with the chosen range of memristance ${R_{OFF}}/{R_{ON}}={100K\Omega}/{1K\Omega}$ values. The result shows the effective conductance reduces as resistance increases when the select transistor is ON. This property helps to take the weighted summation of the sensor out from different cells and hence MAC operation. When the transistor is turned OFF, the extremely small transistor conductance $g_{T}$ dominates the cell conductance. Therefore, the VL and HL currents comes only from those vertical and horizontal ON cells.

Fig \ref{figure7}(b) shows the variation of effective cell conductance $g$ with varying force in $lbs$. The effective conductance increases with increase in the applied force. The figure shows the variation of $g$ with the memristance $1K\Omega$, $10K\Omega$, $50K\Omega$ and $100K\Omega$. Fig \ref{figure7}(a) and Fig \ref{figure7}(b) shows that the proposed TMS crossbars can show significant variation in output with different memristor values and variation in applied force. Thus sensor conductance, $g_{S}$, play a significant role in TMS crossbar in effective cell conductance computations and neuromorphic computing. For the Braille character recognition system, it is considered that the tactile patch is detected with a force of $20 lbs$.

\vspace{-9pt}
\subsection{Braille Alphabet Implementation}
125 distinct Braille characters including Grade 1 and Grade 2 Braille system \cite{braille123} are implemented with the proposed TMS crossbar array. Grade 1 consists of alphabets (capital and small), numbers punctuation and symbols. The Grade 2 braille system represent a shortened form of common words and sounds. Depending on ON/OFF select line of (D7,D8), the 125 braille characters are divided into different groups. Group1: Alphabet Capital (Grade1, 27 symbols), Group2: Alphabet Small (Grade1, 26 symbols), Group3: Words (Grade2, 46 words) and Group4: Numbers, punctuation \& symbols (Grade1, 26 symbols). The Fusion set is the combination of Grade 1, Grade 2, Grade 3 and Grade 4 braille characters with 125 braille symbols. (D1, D2, D3, D4, D5, D6, D7, D8) values (ON, OFF, OFF, OFF, OFF, OFF, OFF, OFF) and (ON, OFF, OFF, OFF, OFF, OFF, OFF, ON) represents A and a respectively.      

The neural computation in the crossbars are prone to noise. This noise can result from thermal noise  or resistance variability. To test the impact of signal noise on the sensor crossbar, we apply additive Gaussian noise with variances, $\sigma^{2}=0.02$, $\sigma^{2}=0.05$, $\sigma^{2}=0.1$ and $\sigma^{2}=0.5$. The dataset consists of 5 copies of the 125 different braille symbols. Table \ref{table1} shows the detection accuracy of implementing the Braille dataset with variation of noise. In the neural network model of the implementation, the first dense layer uses the ReLu activation function, while the last layer is the softmax function. The second neural network layer or first 1T1M crossbar size is taken as 6$\times$14 where as size of output neurons is dependent on the size of the selected group. The accuracy in detecting the symbols for each group is presented and compared with fusion dataset. Table \ref{table1} compares the recognition accuracy of analog TMS crossbar with binary TMS crossbars. The results conveys that analog system performs better in noisy environment whereas binary systems are more prone to errors and hence detection accuracy reduces. Table \ref{table2} shows the comparison of the proposed method with other Braille character recognition system in the literature. The proposed method shows improved performance in terms of detection accuracy for both 26 (Group 1) and 37 (37 characters from Group 3) number of Braille characters. The proposed implementation is scalable solution that can implement 125 Braille characters without much degradation in performance. 

\begin{table}
    \centering
    \caption{\small{Recognition Accuracy of the Analog and Binary TMS crossbar based Braille recognition system. Performance accuracy with variation of noise is also presented.}}
    \label{table1}
     \scalebox{0.7}{
    \begin{tabular}{ |p{0.8cm}|p{0.8cm}|p{0.8cm}|p{0.7cm}|p{0.7cm}|p{0.8cm}|p{0.8cm}|p{0.7cm}|p{0.7cm}|}
  \hline
      Groups  & \multicolumn{4}{|c}{Analog TMS} & \multicolumn{4}{|c|}{Binary TMS}\\
     \cline{2-9}
 &$\sigma^{2}$=0.02&$\sigma^{2}$=0.05&$\sigma^{2}$=0.1&$\sigma^{2}$=0.5 &$\sigma^{2}$=0.02&$\sigma^{2}$=0.05&$\sigma^{2}$=0.1&$\sigma^{2}$=0.5\\
        \hline 
        Group1 &100&98.2&91&85.0 &91.3&88.7&85.6&82.0 \\
        Group2  &96.3&89.0&79&75.2 &94.1&91.0&88.1&85.7\\
        Group3  &97&96.5&87.2&72.6 &85.0&84.4&77.3&72.6\\
        Group4 &  97.8&94.1&90.0&82.5 &  94.5&94.0&85.0&72.2\\
        \hline
   Fusion  &  96.0&92.1&89.9&87 &  85.2&83.0&82.2&79.0 \\
        \hline
       \end{tabular}}
    \end{table}

\begin{table}[]
    \centering
    \caption{\small{Braille Character Recognition-A Comparison. Convolutional Neural Network (CNN), Deep Convolutional Neural Network (DCNN), Artificial Neural Network (ANN)}}
    \label{table2}
    \scalebox{0.7}{
    \begin{tabular}{ |p{3cm}|p{2.3cm}|p{2.7cm}|p{1.5cm}|}
  \hline
      Methodology  & Neural Architecture & No of Braille characters & Accuracy(\%)\\
     \hline
      Braille Image \cite{sym12071069} &  CNN &37&98.73\\
      Binary encoding \cite{sym0001} & 3 layer ANN & 26  & 99\\
      Braille Image \cite{braille444, 9152576} & CNN & 26 & 95.8, 94\\
      Braille Image\cite{cmc.2021.015614}& DCNN & 26 / 37 & 99 / 98\\
      Proposed  & 3 layer ANN &26 / 37 / 125 &100 / 98 / 96 \\
        \hline
       \end{tabular}}
    \end{table}
    
\begin{figure}
	\centering
	\subfloat[]{\includegraphics[height=1.1in,width=1.4in]{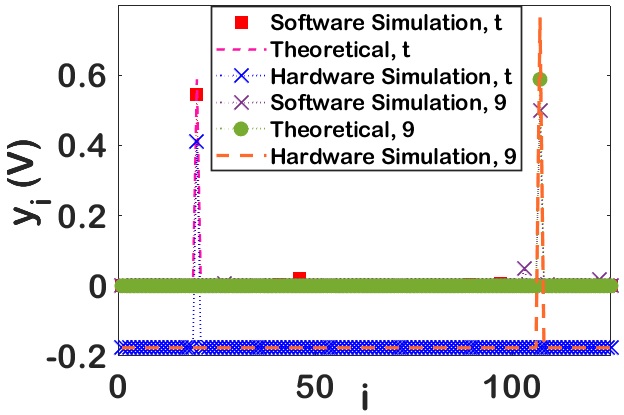}}\quad
	\subfloat[]{\includegraphics[width=1.4in]{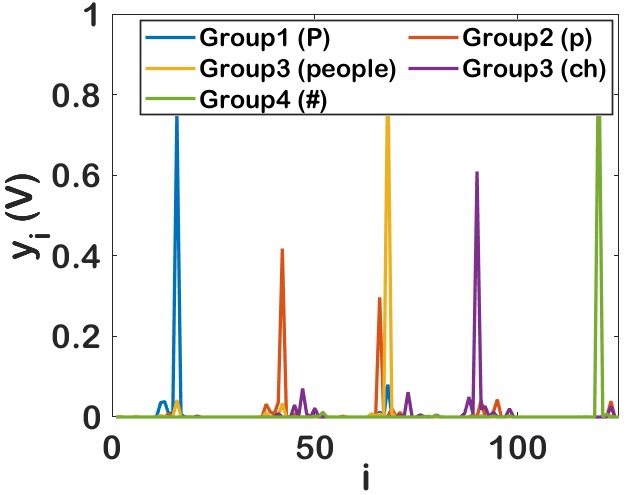}}\\
	\subfloat[]{\includegraphics[width=1.4in]{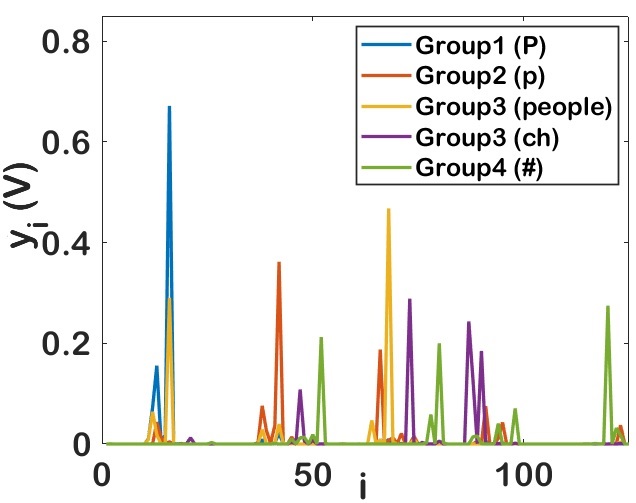}}\quad
	\subfloat[]{\includegraphics[width=1.4in]{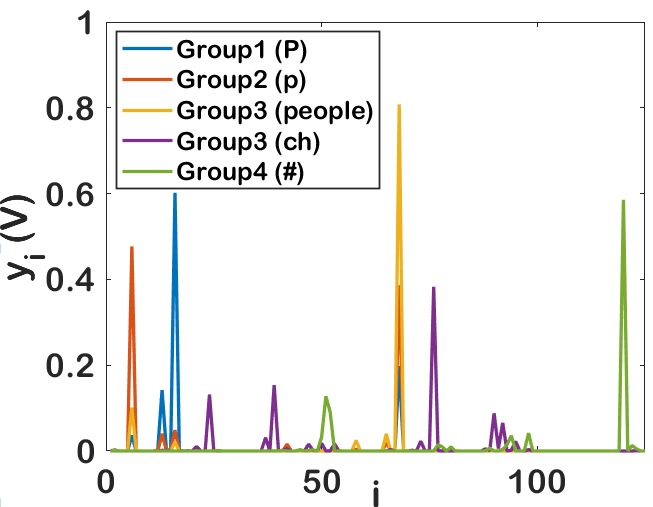}}\\
	\caption{\small{Output of Braille recognition system (a) Comparison of Software, Hardware and Theoretical output (b) Without noise (c) Noise with $\sigma^2=0.05$ (d) Noise with $\sigma^2=0.1$}}
	\label{figure71}
\end{figure}

The trained weights are then mapped to the memristor crossbar array. In the softmax circuit, $I_{s}=1nA$ and $R_{f}=100K\Omega$. The output of the softmax activation function of the proposed system is shown in Fig \ref{figure71}(a) for Group1, alphabet 't' and Group4, number '9'. The softmax activation function represents the output with probabilities and the output port with maximum probability is the detected symbol. According to the figure Group1, alphabet 't' refers to port 20 and  Group4, number '9' refers to port 107. The output is validated with the theoretical value in eqn. (\ref{theoretical}) and the software simulated output with $ \left ( \frac{e^{a_{z}}}{\sum_{i}e^{a_{z}} } \right ) $  using Python programming. For hardware out, softmax output calculated by taking the exponential of  $\frac{V_{m}}{V_{T}}$. Hence the minimum probable symbol values approximates to zero and maximum probable symbol values will be enhanced.

Fig \ref{figure71}(b), Fig \ref{figure71}(c) and Fig \ref{figure71}(d) shows the variation in softmax output with noise. The output of the proposed design tested with Group1- alphabet 'P', Group2 - alphabet 'p', Group3 - word 'people', Group3 - sound 'ch', Group4 - symbol '\#' is presented. Fig \ref{figure71}(b) shows the output without noise and the symbol is correctly identified at port 16, 42, 68, 90 and 120 respectively. If the same input is corrupted with noise of variance $\sigma^{2}=0.05$, 'P', 'p', 'people' and '\#' are detected correctly but 'ch' will be falsely detected as 'us'. Whereas with noise of variance $\sigma^{2}=0.05$, 'p' and 'ch' will be falsely detected as 'f' and 'it' respectively. This shows that as noise increases false detection increases and reduces accuracy.     

\vspace{-9pt}
\subsection{Area and Power Calculation}
{Table \ref{table3} provides the area and power requirements of the proposed TMS-based Braille system. The power and area requirements of crossbars and amplifiers are listed separately in Table \ref{table3}. The  output of crossbars can be parallel processed or serial processed \cite{8468181}. A detailed description of serial processing in TMS crossbars is presented in Supplementary Material. Since the power consumption of parallel processing increases to 0.2W, the serial processing of with an analog signal multiplexing and a single division circuit, with power consumption of 3.7mW, can be considered a power efficient way of implementing the TMS-based Braille system. With the sensor area of $50.8mm \times 50.8mm$ \cite{flxa505}, the area of Layer 1 rises to 0.02$m^{2}$ whereas the other layers gives an area of 512.898E-6 $m^{2}$ for parallel processing and 129.6E-6 $m^{2}$ for serial processing respectively. Table \ref{table3} also presents the power and area calculation of the Binary TMS implementation of Braille system for parallel and series processing as in \cite{8626224}. The sensors give analog output, and the tactile solutions implemented as binary systems require data converters, interface, and adder circuits that require higher area and power than analog circuits.}

\begin{table}[]
 \centering
  \caption{\small{Power Consumption and Area of TMS-based Braille Character Recognition System: A comparison between Series and Parallel processing}}
    \label{table3}
\scalebox{0.6}{
\begin{tabular}{|p{2.8cm}|p{1.3cm}|p{0.7cm}|p{1.3cm}|p{0.7cm}|p{1.3cm}|p{0.7cm}|p{1.3cm}|p{0.7cm}|}
\hline
\multirow{2}{*}{\textbf{Circuit block}} & \multicolumn{2}{l|}{\textbf{Analog Parallel}}       & \multicolumn{2}{l|}{\textbf{Analog Series}}   & \multicolumn{2}{l|}{\textbf{Binary Parallel}}       & \multicolumn{2}{l|}{\textbf{Binary Series}}        \\ \cline{2-9} 
                                        & \textbf{Area ($m^{2}$)} & \textbf{Power (W)}        & \textbf{Area ($m^{2}$)} & \textbf{Power (W)} & \textbf{Area ($m^{2}$)} & \textbf{Power (W)}       & \textbf{Area ($m^{2}$)} & \textbf{Power (W)}        \\ \hline
\textbf{TMS Crossbar (layer1)}                & 0.02                    & \multirow{2}{*}{262$\mu$} & 0.02                    & \multirow{2}{*}{262$\mu$} & 0.02                    & \multirow{2}{*}{3.6$m$} & 0.02                    & \multirow{2}{*}{3.6$m$} \\ \cline{1-2} \cline{4-4} \cline{6-6} \cline{8-8}
\textbf{Crossbar (layer 2 \& 3)}             & 39.5E-6             &                           & 39.5E-6             &          & 562.3E-6                &                           & 562.3E-6                 &                           \\ \hline
\textbf{Amplifiers (layer 1)}           & 3.38E-6                 & \multirow{2}{*}{236.4m}   & 0.906E-6                 & \multirow{2}{*}{3.4m}  & 6.77E-6               & \multirow{2}{*}{1.9}   & 1.47E-6               & \multirow{2}{*}{0.1}      \\ \cline{1-2} \cline{4-4} \cline{6-6} \cline{8-8}
\textbf{Amplifiers (layer 2\&3)}                    & 438.45E-6              &                           & 90E-6               &    & 2932E-6                  &                                             & 154E-6                  &                           \\ \hline
\end{tabular}}
\end{table}

\vspace{-10pt}
\section{Conclusion}
The paper demonstrated a TMS crossbar array that integrates the near-sensor  neural computation for tactile sensing. The paper presents how the conventional one transistor crossbar can be modified to enable both row and column readouts to improve the scalability of the proposed approach. The results shows that the effective conductance of each cell is directly depended on sensor and memristor conductance values. The Braille character recognizing system with the proposed methods shows better accuracy and scalability in comparison with other Braille character recognizing system in the literature. The proposed TMS crossbar array exhibits the ability of transforming a tactile sensor path to a system with sensing and analog neural computing capabilities. The discussion on benefits of analog implementation of the proposed method is substantiated with proper results in terms of accuracy, area and power requirements in comparison with the binary counterparts. In addition to superior accuracy, proposed design reduces use of data converters, interface circuits and transmission modules.



%





\ifCLASSOPTIONcaptionsoff
  \newpage
\fi

\end{document}